\documentclass[10pt,a4paper]{article}
\pdfoutput=1
\usepackage[utf8]{inputenc}
\usepackage{verbatim}
\usepackage{amsthm}
\usepackage{amsmath}
\usepackage{amssymb}
\usepackage{graphicx, subfigure}
\usepackage{esint}

\usepackage{placeins}
\usepackage{siunitx}

\usepackage{jcappub}
\newcommand\Tstrut{\rule{0pt}{2.6ex}} 

\begin{document}

\title{Constraining dynamical neutrino mass generation with cosmological data}

\author[a]{S. M. Koksbang,}
\author[a]{S. Hannestad}
\affiliation[a]{Department of Physics and Astronomy \\
University of Aarhus, DK-8000 Aarhus C, Denmark}

\emailAdd{sth@phys.au.dk}

\date{\today}

\abstract{We study models in which neutrino masses are generated dynamically at cosmologically late times. Our study is purely phenomenological and parameterized in terms of three effective parameters characterizing the redshift of mass generation, the width of the transition region, and the present day neutrino mass. We also study the possibility that neutrinos become strongly self-interacting at the time where the mass is generated. We find that in a number of cases, models with large present day neutrino masses are allowed by current CMB, BAO and supernova data. The increase in the allowed mass range makes it possible that a non-zero neutrino mass could be measured in direct detection experiments such as KATRIN.
Intriguingly we also find that there are allowed models in which neutrinos become strongly self-interacting around the epoch of recombination.}

\maketitle

\section{Introduction}
One of the major discoveries in particle physics in the last two decades is the presence of non-zero neutrino masses, measured through neutrino oscillations in experiments such as Super-Kamiokande \cite{kamiokande}, solar SNO \cite{solarSNO} and KamLAND \cite{KamLAND}. These oscillations require at least two of the three neutrino mass eigenstates to be massive. However, neutrinos are much lighter than other fermions in the standard model. Data from measurements of neutrino oscillations have established the presence of two mass differences of order $\Delta m_{21}^2 \sim 7.5 \times 10^{-5} \, {\rm eV}^2$ and  $\Delta m_{3(1/2)}^2 \sim 2.5 \times 10^{-3} \, {\rm eV}^2$ respectively (see e.g.\ \cite{Esteban:2016qun}). Together with direct measurements from tritium decay (see e.g.\ \cite{Kraus:2004zw}) this constrains the masses of all active neutrinos to be below the eV scale. This is quite puzzling since neutrino masses must then be many orders of magnitude below what might naturally be expected of Dirac fermions coupling to the Higgs.

This puzzle has led to intense theoretical work on neutrino mass generation mechanisms (see {\em e.g.} \cite{fukugita_yanagida} for a thorough review). The most widely studied model is the see-saw mechanism which relies on the possibility that neutrinos can have very large Majorana masses (see e.g. \cite{mass_review,seesaw_study}). Combined with Dirac masses of the same order of magnitude as for other fermions, this leads to a separation of scales with three light left handed and three heavy right handed Majorana neutrinos.
\newline\indent 
Especially since there is no known way to test the see-saw mechanism, it is important to also study other possible methods of neutrino mass generation. Some suggested alternative mechanisms provide dynamical generations of neutrino masses on cosmologically interesting timescales.
For instance, in the mass varying neutrinos (MaVaN) scenario \cite{Fardon:2003eh}, neutrinos couple to a light scalar field which slowly rolls in a flat potential. This leads to the interesting property that neutrino masses are directly linked to dark energy which would explain why the neutrino mass scale is comparable to the energy scale required for quintessence-like dark energy ($E \sim 10^{-3}$ eV). Studies (see {\em e.g.} \cite{Afshordi:2005ym}) have however shown that because the coupling to a light scalar mediates an attractive force between neutrinos, these quickly form bound structures. While this is not a problem from a structure formation point of view, it means that the light scalar field cannot explain dark energy, except under rather special circumstances.
\newline\indent
Another, more recent, attempt at explaining dynamical neutrino mass generation is the work of Dvali and Funcke \cite{dvali}. There, neutrino masses are generated by a gravitational $\theta$-term at a late time phase transition (at $T \sim m_\nu$). The model predicts that the onset of a non-zero neutrino mass at the phase transition temperature is rapid and that the resulting fluid is strongly self-interacting. Similar phenomenology is obtained with the models studied in {\em e.g.} \cite{similar_to_dvali,a_little_similar}.
\newline\newline
In this paper, we study cosmological constraints on dynamical neutrino mass generation from a phenomenological point of view, inspired by the qualities of the model in \cite{dvali}. In particular, we will assume that neutrinos are massless until a specific value of the scale factor is reached. At this time the mass increases to the present-day value and remains constant. We additionally include the possibility that neutrinos become strongly self-interacting at the time of mass generation.
\newline\indent
We will allow for a finite width of the mass-transition, leading to two parameters that control the epoch and duration of the transition. We have implemented the dynamical neutrino mass model in the Boltzmann code CAMB \cite{Lewis:1999bs} in order to test the effect on cosmological observables, and used CosmoMC \cite{Lewis:2002ah} to constrain the models with CMB, BAO and supernova data. The dynamical neutrino mass model was also implemented into CLASS \cite{Blas:2011rf} in order to double-check the implementation in CAMB.
\newline\newline
The paper is structured as follows. In section \ref{sec:Theory} we briefly review the Boltzmann equations and discuss the effects of introducing our non-standard neutrino phenomenology. Section \ref{sec:data} describes the models and data used in the work while results are presented and discussed in section \ref{sec:Results}. We conclude in section \ref{sec:Conclusion}.

\section{The Boltzmann equations}\label{sec:Theory}
The first part of this section serves to give a brief summary of the evolution equations of neutrinos according to first order perturbation theory. For further details on this matter, the reader is referred to {\em e.g.} \cite{MaAndBert} which is the basis for the summary given below. In following subsections, the effects of neutrino mass and strong neutrino self-interactions are discussed. 
\newline\newline
The background metric is assumed to be a flat FLRW metric and perturbations are given in the conformal Newtonian gauge. The appropriate line element is therefore
\begin{equation}
ds^2 = a^2(\tau)\left( -\left(1+2\psi \right) d\tau^2 + \left( 1-2\phi\right) dx^idx_i   \right) ,
\end{equation}
where the standard convention of setting $c = 1$ is used.
\newline\newline
The FLRW metric requires neutrinos to be described as a perfect fluid at background order. At first order, however, a perfect fluid approximation is generally not an adequate description for neutrinos. Instead, neutrinos must be described through their phase space distribution function $f = f(x^i, P_j, \tau)$, where $\tau$ is proper time and $P^\mu = m\frac{dx^{\mu}}{d\tau}$ is the conjugate momentum. In the Newtonian gauge, the conjugate momentum is related to the proper momentum, $p^i$, by $P_i = a\left(1-\phi \right)p_i  $. At background level, the distribution function is simply the Fermi-Dirac distribution, {\em i.e.} $\bar f \propto \frac{1}{e^{\epsilon/\left(aT_{\nu} \right)} +1}$. A bar is used to denote background quantities while $\epsilon = a^2\sqrt{p^2+m^2} = \sqrt{q^2+m^2a^2}$ will be used below to denote the neutrino energy.
\newline\newline
The evolution of the neutrino phase space distribution is given by the relativistic Boltzmann equation which simply describes neutrino number conservation. Assuming, for the moment, that neutrinos are collisionless, this reduces to the relativistic Vlasov equation,
\begin{equation}
\frac{df}{dt} = P^\mu \frac{\partial f}{\partial x^\mu}-\Gamma^i_{\mu\nu}P^{\mu}P^\nu\frac{\partial f}{\partial P^i} = 0.
\end{equation}
Writing $f$ in terms of its background part, $\bar{f}\left( q\right) $, and perturbation, $\Psi\left(x^\mu, q, n_j \right) $, where $\vec{q}^j =qn^j $, the Fourier transform of the Vlasov equation can be written as
\begin{equation}
\frac{\partial \Psi}{\partial \tau} + i\frac{q}{\epsilon}\left( \vec{k}\cdot \hat{n}\right)\Psi + \frac{d\ln \bar{f}}{d\ln q}\left[ \dot \phi -i\frac{\epsilon}{q}\left(\vec{k}\cdot \hat{n} \right)\psi \right] = 0    ,
\end{equation} 
where a dot is used to denote partial derivatives with respect to conformal time.
\newline\indent
By expanding the perturbation $\Psi(\vec k,\hat n, q, \tau )$ in a Legendre series as 
\begin{equation}
\Psi = \sum\limits_{l = 0}^{\infty}\left( -i\right) ^l\left( 2l+1\right) \Psi_l\left( \vec k,q,\tau\right)  P_l\left( \hat k \cdot \hat n\right) ,
\end{equation}
the first order part of the Vlasov equation reduces to the infinite hierarchy

\begin{equation}\label{eq:hierarchy}
\begin{split}
\dot \Psi_0 = -\frac{qk}{\epsilon}\Psi_1 -\dot \phi \frac{d\ln \bar{f}}{d\ln q}\\
\dot \Psi_1 = \frac{qk}{3\epsilon}\left(\Psi_0 -2\Psi_2 \right) - \frac{\epsilon k}{3q}\psi \frac{d\ln \bar{f}}{d\ln q}\\
\dot \Psi_l = \frac{qk}{\left(2l+1 \right) \epsilon}\left[l\Psi_{l-1} -\left( l+1\right) \Psi_{l+1} \right] ,\,\,\,\,\, l\leq 2  
\end{split}
\end{equation}
The neutrinos are related to the other components of the Universe and to the metric perturbations through their stress-energy-momentum tensor which is obtained from their distribution function by the relation
\begin{equation}
T_{\mu\nu} = \int d^3 P \frac{1}{\sqrt{-g}}\frac{P_{\mu}P_\nu}{P^0}f\left(x^i, P_j,\tau \right) .
\end{equation}
In particular, the energy density perturbation, velocity divergence and shear of the neutrino stress-energy-momentum tensor are defined as
\begin{equation}
\begin{split}
\delta:=\frac{\delta\rho}{\bar\rho} = \frac{\delta T^0_0}{\bar T^0_0}\\
\theta:= \frac{ik_j\delta T^0_j}{\bar \rho + \bar p}\\
\sigma:= -\left( \hat k_i\hat k_j -\frac{1}{3}\delta_{ij}\right) \left( T^{ij} -\frac{1}{3}\delta^{ij}T^k_k\right) ,
\end{split}
\end{equation}
where $p = \omega\rho$.
\newline\newline
With these definitions, the integrals of the first two equations of the Boltzmann hierarchy in equation (\ref{eq:hierarchy}) over $q^2\epsilon dq$ and $q^3dq$, respectively, yields (see {\em e.g.} \cite{heirarchy_to_delta}) the two evolution equations
\begin{equation}\label{eq:delta}
\begin{split}
\dot \delta = -\left( 1+\omega\right) \left( \theta -3\dot{\phi}\right) -3\mathcal{H}\left( \frac{\delta P}{\delta \rho} -\omega\right)\delta 
\\
\dot \theta = -\mathcal{H}\left( 1-3\omega\right) \theta -\frac{\dot \omega}{1+\omega}\theta +\frac{\delta P/\delta\rho}{1+\omega}k^2\delta -k^2\sigma + k^2\psi
,
\end{split}
\end{equation}
where $\mathcal{H}$ is the conformal Hubble parameter.
\newline\newline
The following subsections discuss how the introduction of neutrino masses and self-interactions are implemented and affect cosmological observations.

\begin{figure}[htb!]
\centering
\includegraphics[scale = 1.2]{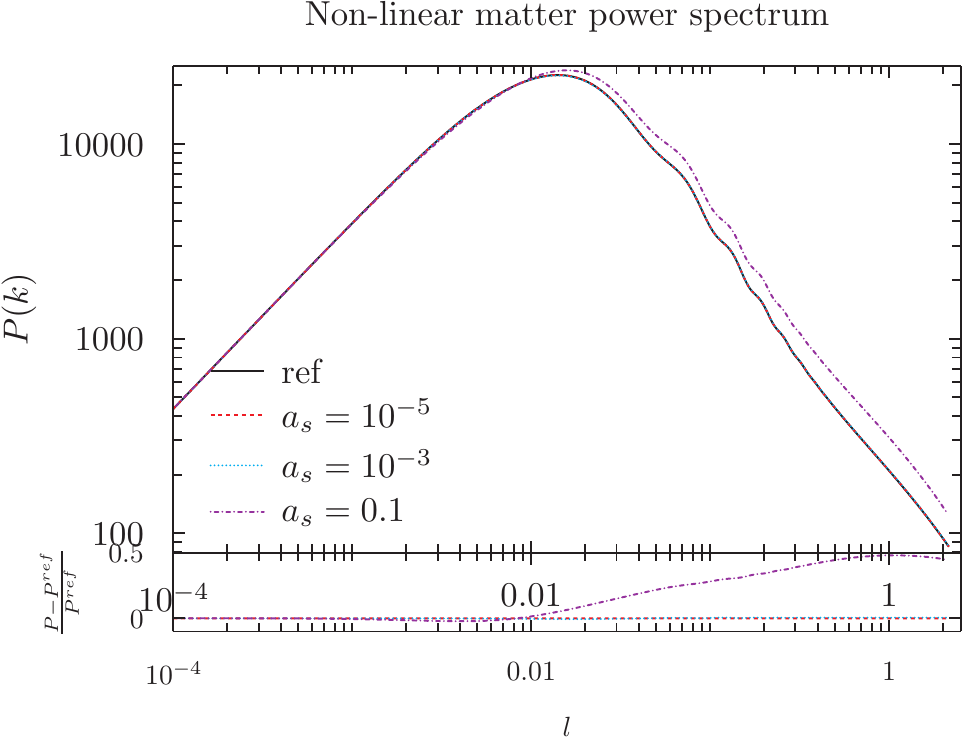}
\caption{The power spectrum for variable values of $a_{s}$ with all other parameters kept fixed. Neutrino masses are set to $m_{\nu,0} = 0.31$eV. $B_{s}$ is set to $10^{10}$ in all models, approximating instantaneous mass transition. The reference model is defined by $a_{s} = 10^{-13}$, corresponding to a model with fixed neutrino mass.}
\label{fig:power_mass}
\end{figure}

\subsection{Effects of neutrino masses}
The neutrinos are modeled as being massless until a specific time, $a = a_{s}$. When this scale factor is reached, the neutrino mass increases to its present day value $m_{\nu}$. More precisely, the neutrino mass is modeled using the following prescription:
\begin{equation}\label{eq:mass}
m_\nu (a) = \left\{ \begin{array}{rl}
0  &\text{if} \,\, a<a_s \\
m_{\nu}\tanh\left(B_{s} \left[\frac{a}{a_{s}} -1 \right] \right) &\text{if} \,\, a\geq a_s,
\end{array} \right.
\end{equation}
where $m_{\nu}$ is the present time neutrino mass and $B_{s}$ is a free parameter which determines the rate of transition between $m_\nu(a) = 0$ and $m_\nu(a) = m_{\nu}$.
We note here that the specific form of Eq.~\ref{eq:mass} is not crucial. We have tested several other forms and found them to give similar results.
\newline\newline
If $a_{s}$ is very small, say $10^{-13}$, the neutrinos are effectively massive at all times and the situation reduces to the standard massive-neutrinos picture. If $B_{s}$ is very large, say $10^{10}$, the transition between $m_{\nu}(a) = 0$ and $m_{\nu}(a) = m_{\nu}$ can be considered instantaneous. This combination of values of $a_s$ and $B_s$ will be used as the reference case.
\newline\indent
The effect of turning on neutrino masses depends on the size of $m_{\nu}$ and in particular on whether the neutrinos are still light enough to be considered radiation. For simplicity, the following discussion applies to $m_{\nu}$ small enough for the neutrinos still to be considered as radiation even when massive (except at very late times).
\newline\newline
\begin{figure}[htb!]
\centering
\includegraphics[scale = 1.2]{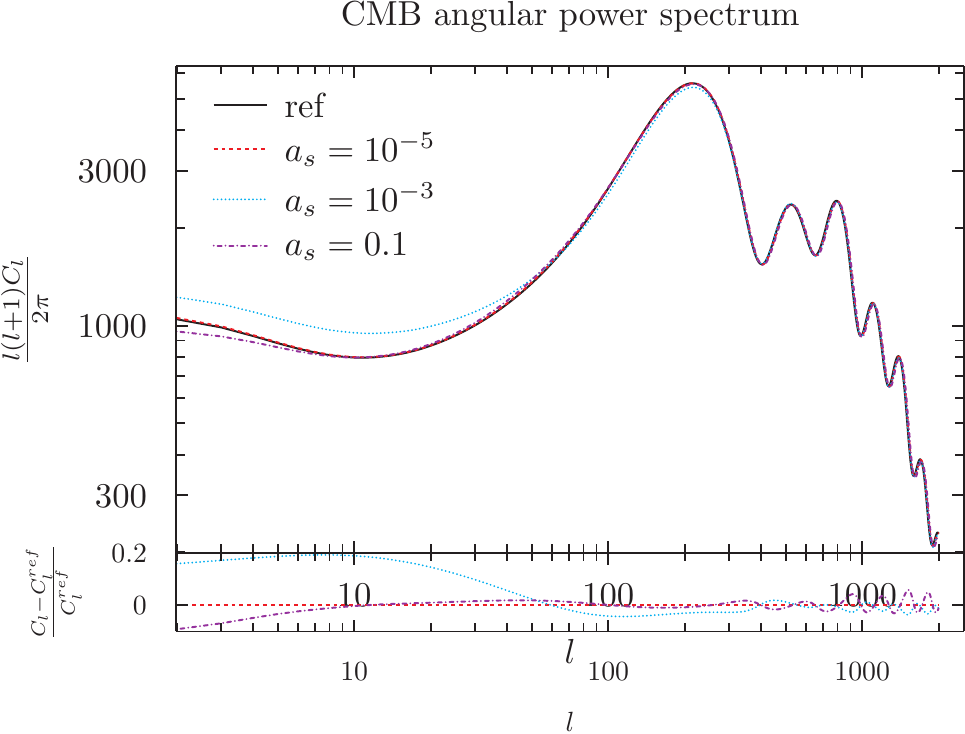}
\caption{The CMB temperature anisotropy spectrum for variable values of $a_{s}$, with all other parameters kept fixed. Neutrino masses are set to $m_{\nu} = 0.31$eV. $B_{s}$ is set to $10^{10}$ in all models, corresponding to instantaneous mass transition. The reference model is defined by $a_{s} = 10^{-13}$ corresponding to a model with fixed neutrino mass. Other parameters are set in accordance with the standard $\Lambda$CDM model, {\em i.e.} $H_0 \approx 70$km/s/Mpc, $\Omega_{c} \approx 0.3$ and $\Omega_{\Lambda}\approx 0.7$.}
\label{fig:cmb_mass}
\end{figure}
\begin{figure}[!htb]
\centering
\subfigure[]{
\includegraphics[scale = 0.7]{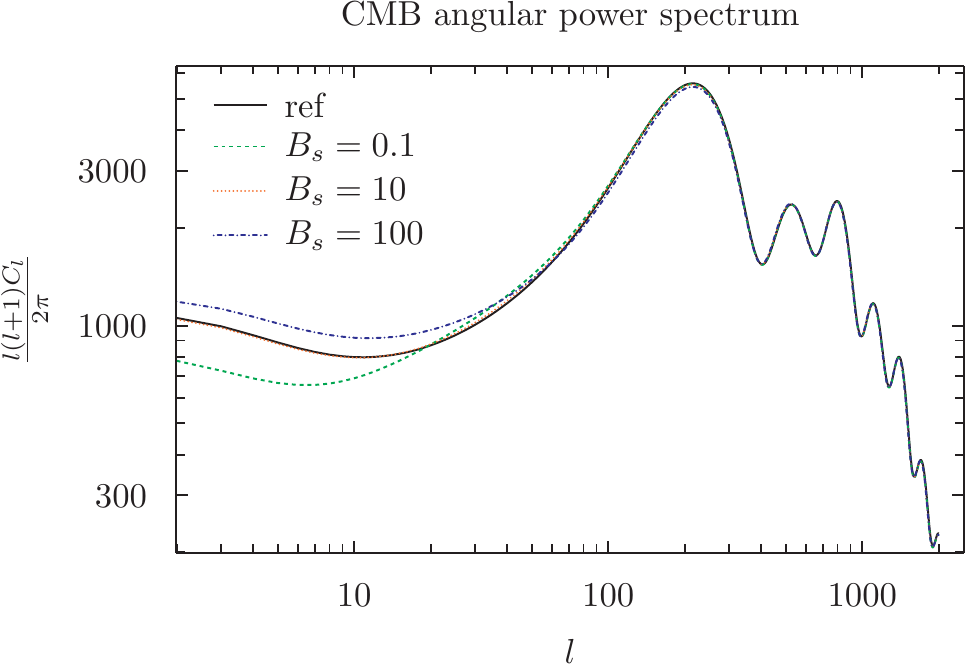}
\label{subfig:a}
}
\subfigure[]{
\includegraphics[scale = 0.7]{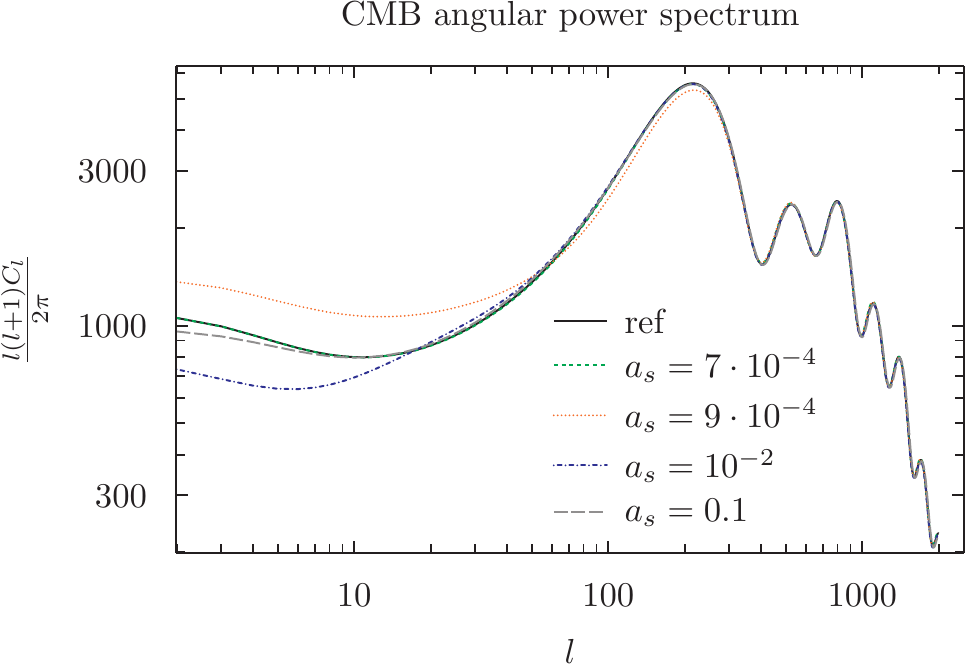}
\label{subfig:b}
}
\caption{The CMB temperature anisotropy spectrum for standard models augmented with mass of neutrinos switched on at $a = a_{s}$. The figure to the left compares the spectra for different values of $B_{s}$ for fixed $a_{s} = 10^{-3}$ while the figure to the right compares the spectra for different values of $a_{s}$ with fixed $B_{s} = 10^{10}$. Other parameters are set in accordance with the standard $\Lambda$CDM model, {\em i.e.} $H_0 \approx 70$km/s/Mpc, $\Omega_{c} \approx 0.3$ and $\Omega_{\Lambda}\approx 0.7$.}
\label{fig:cmb_special}
\end{figure}
While other components of the Universe cluster to form structures, neutrinos free-stream. As discussed in {\em e.g.} \cite{Pastor}, this free-streaming reduces the growth of other components' perturbations because neutrinos contribute to the homogeneous expansion of space (causing the Newtonian metric potentials to decay) without contributing to the gravitational clustering on scales below the neutrinos' free-streaming scale  (which is mass dependent). The suppression increases as the neutrino masses increase since a larger neutrino mass increases the overall neutrino contribution to cosmological evolution. Hence, in the models studied here, the power spectrum will be suppressed more, the earlier the neutrino masses are turned on and the larger $m_\nu$ is. This is illustrated in figure \ref{fig:power_mass} which shows the power spectrum for models of different values of $a_{s}$, all with $B_{s} = 10^{10}$. Note that the power spectra have slightly different turn-over scales depending on whether the neutrinos become massive before or after radiation-matter equality. With the low neutrino masses studied here, neutrinos are relativistic during radiation-matter equality and will thus contribute to the radiation energy of the Universe during this epoch. Therefore, models where neutrino masses are turned on before radiation-matter equality will have a slightly delayed radiation-matter equality. This in turn implies a suppression of the power spectrum up to slightly larger scales since these will enter the horizon before radiation-matter equality and hence be decaying until the Universe becomes matter-dominated. Besides affecting the turn-over scale, the delayed radiation-matter equality also leads to an overall suppression of structure formation on small scales.
\newline\newline
\begin{figure}[!htb]
\centering
\subfigure[]{
\includegraphics[scale = 0.7]{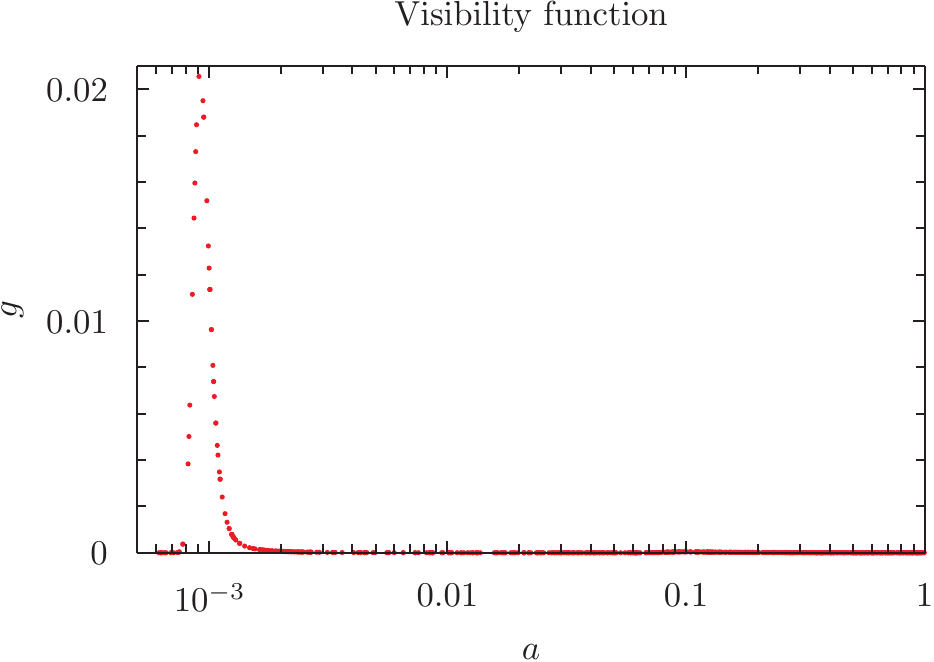}
}
\subfigure[]{
\includegraphics[scale = 0.7]{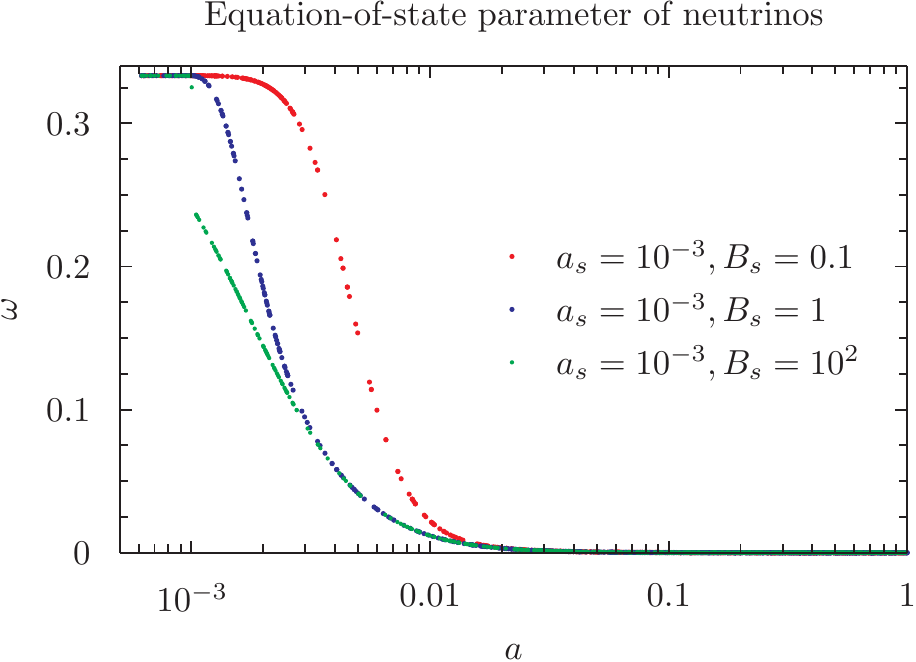}
}
\caption{Visibility function, $g(\tau)$, and equation of state parameter for $a_s = 10^{-3}$. The plotted relations were obtained with CAMB.}
\label{fig:g_omega}
\end{figure}
In principle, the generation of neutrino masses will also affect CMB anisotropies. For instance, a delayed radiation-matter equality induces a larger early Integrated Sachs-Wolfe (eISW) effect which adds in phase to the Sachs-Wolfe effect and hence amplifies the CMB anisotropies around the first peak. The eISW effect is only important on scales large enough for a structure to evolve significantly during the time it takes a light ray to traverse it.
\newline\indent
As seen in figure \ref{fig:cmb_mass}, the overall effect of $a_{s}$ for a fixed $m_{\nu}$ on the CMB angular power spectrum is quite modest except at a certain interval around $a_{s} = 10^{-3}$. When the neutrino mass is turned on close to the time of last scattering, a quite significant effect is seen in the low-$l$ part of the CMB spectrum. This can be understood by looking at the expression for the CMB temperature multipoles, $\Theta_l$, (see {\em e.g.} equation 8.55 in \cite{dodelson}):
\begin{equation}\label{eq:dodelson}
\begin{split}
\Theta_l(k,\tau_0) = \int_{0}^{\tau_0}d\tau g(\tau)\left( \Theta_0(k,\tau) + \Psi(k,\tau) \right) j_l\left[k(\tau_0-\tau) \right]\\
- \int_{0}^{\tau_0}d\tau g(\tau) \frac{iv_b}{k}\frac{d}{d\tau}j_l\left[k(\tau_0-\tau) \right]\\
+ \int_{0}^{\tau_0} d\tau e^{-\tau}\left[ \dot \Psi - \dot \Phi\right] j_l\left[k(\tau_0-\tau) \right]
\end{split}
\end{equation}
The effects seen in the CMB angular power spectrum in figure \ref{fig:cmb_special} are due to two different physical effects. First of all, when the mass is turned on, the equation of state parameter of neutrinos changes quite abruptly, leading to a large value of $\dot\omega$ which again affects $\delta$ (as seen by equation \ref{eq:delta}) and hence the CMB angular power spectrum. This leads to prominent changes in the CMB angular power spectrum through the first two lines in equation (\ref{eq:dodelson}), but only if it happens around the time of last scattering; the two top lines in equation (\ref{eq:dodelson}) are multiplied by the visibility function, $g(\tau)$, which is narrowly peaked about the time of last scattering as seen in figure \ref{fig:g_omega}. Notice also that the faster the neutrino mass goes from zero to $m_{\nu}$, the larger $\dot\omega$ is (see figure \ref{fig:g_omega}) and hence varying $B_s$ should lead to visible effects in the CMB angular power spectrum for $a_s\sim 10^{-3}$. This is illustrated in figure \ref{subfig:a}.
\newline\indent
\begin{figure}[!htb]
\centering
\includegraphics[scale = 1]{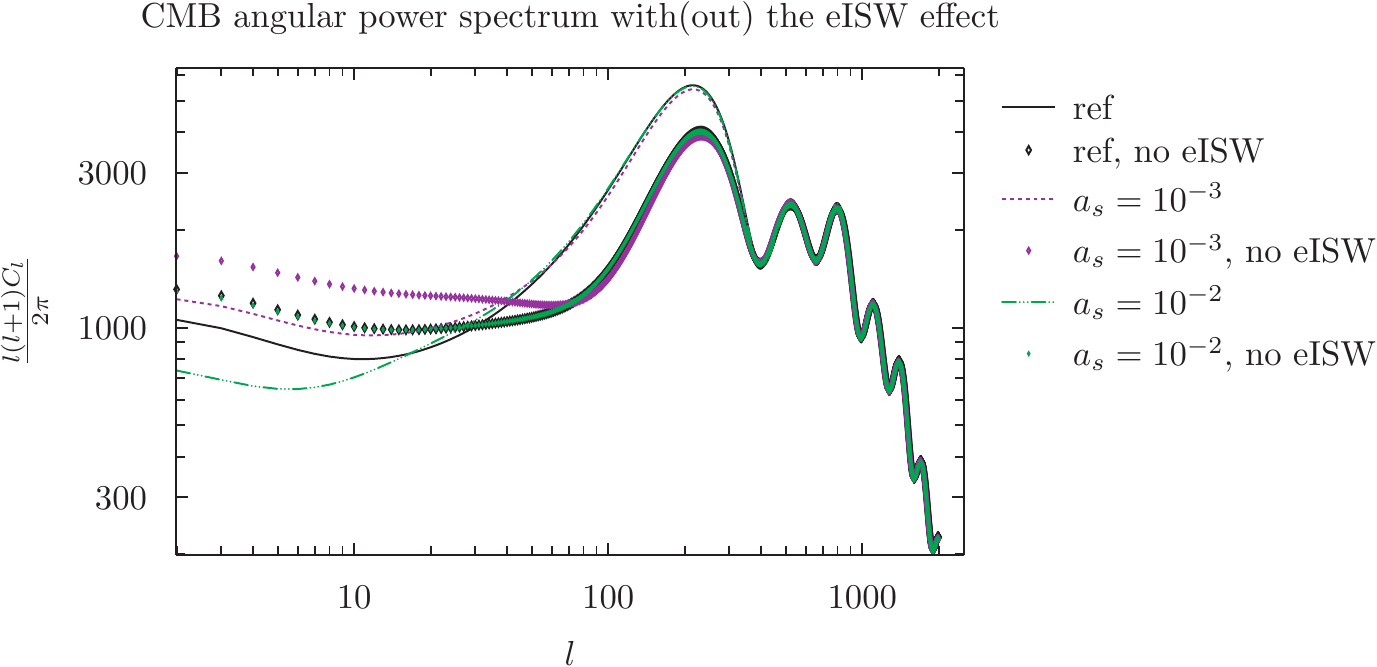}
\caption{Angular power spectrum of the CMB with and without the early ISW effect.}
\label{fig:eISW}
\end{figure}
The above consideration explains the prominent alteration of the angular power spectrum of the CMB if $a_s\approx a_{ls}\approx 10^{-3}$. To understand why the CMB also has prominent effects for $a_s$ as high as $\sim 10^{-2}$ (see figure \ref{subfig:b}), the third line in equation (\ref{eq:dodelson}) must be considered. The third line is the ISW effect and depends on the temporal changes in the metric potentials. When the neutrinos gain mass, the potentials change abruptly which leads to a significant eISW effect if the neutrino mass is generated sufficiently close to the time of last scattering. In particular, a neutrino mass generation has to occur early enough for the neutrino energy contribution still to be a significant part of the total energy density of the Universe. This is seen in figure \ref{fig:eISW} where the CMB angular power spectrum is shown for three values of $a_s$ with and without the eISW effect. More precisely, figure \ref{fig:eISW} shows the CMB angular power spectrum with and without the eISW effect included, for three different models, one being the reference model and the others having $a_s = 10^{-2}$ and $a_s= 10^{-3}$. The figure shows that the CMB angular power spectrum for the $a_s = 10^{-2}$ case is virtually the same as that of the reference model if the eISW effect is removed. On the other hand, the figure shows that for the $a_s = 10^{-3}$ case, removing the eISW effect still leaves a significant deviation from the reference model's CMB angular power spectrum.

\FloatBarrier

\subsection{Effects of neutrino self-interactions}
\begin{figure}[htb!]
\centering
\includegraphics[scale = 1.2]{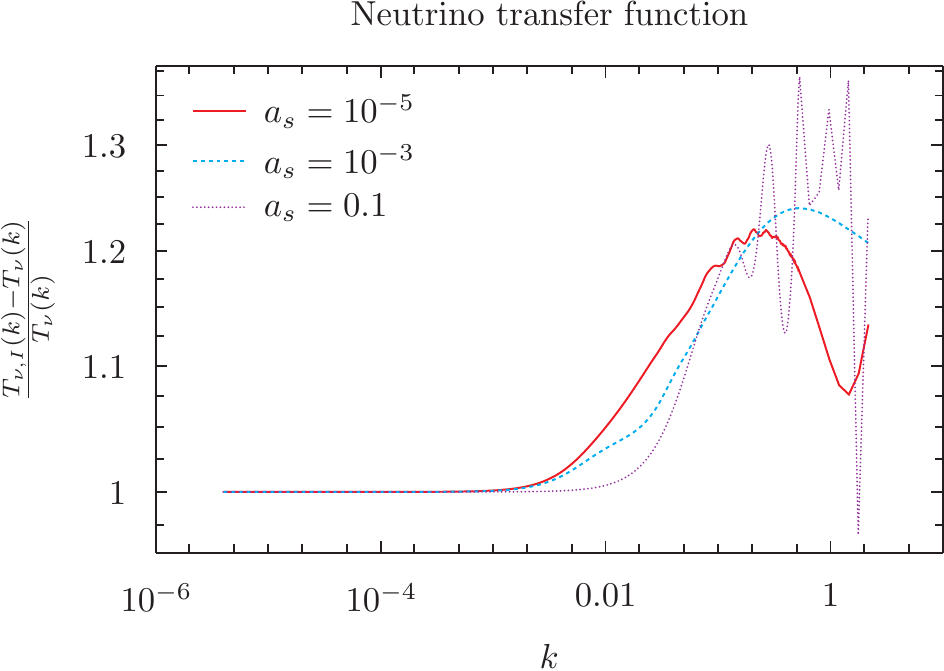}
\caption{Neutrino transfer function comparison for models with and without neutrino self-interactions after $a = a_{s}$, for variable values of $a_{s}$ and all other parameters kept fixed with $m_{\nu} = 0.31$eV and $B_{s} = 10^{10}$ and other parameters set in accordance with the standard $\Lambda$CDM model, {\em i.e.} $H_0 \approx 70$km/s/Mpc, $\Omega_{c} \approx 0.3$ and $\Omega_{\Lambda}\approx 0.7$. $T_{\nu}$ denotes the neutrino transfer function for a given model {\em without} self-interactions while $T_{\nu,I}$ denotes the same for the corresponding model {\em with} strong interactions for $a \geq a_s$.}
\label{fig:transfer_nu}
\end{figure}
\begin{figure}[!htb]
\centering
\subfigure[]{
\includegraphics[scale = 0.7]{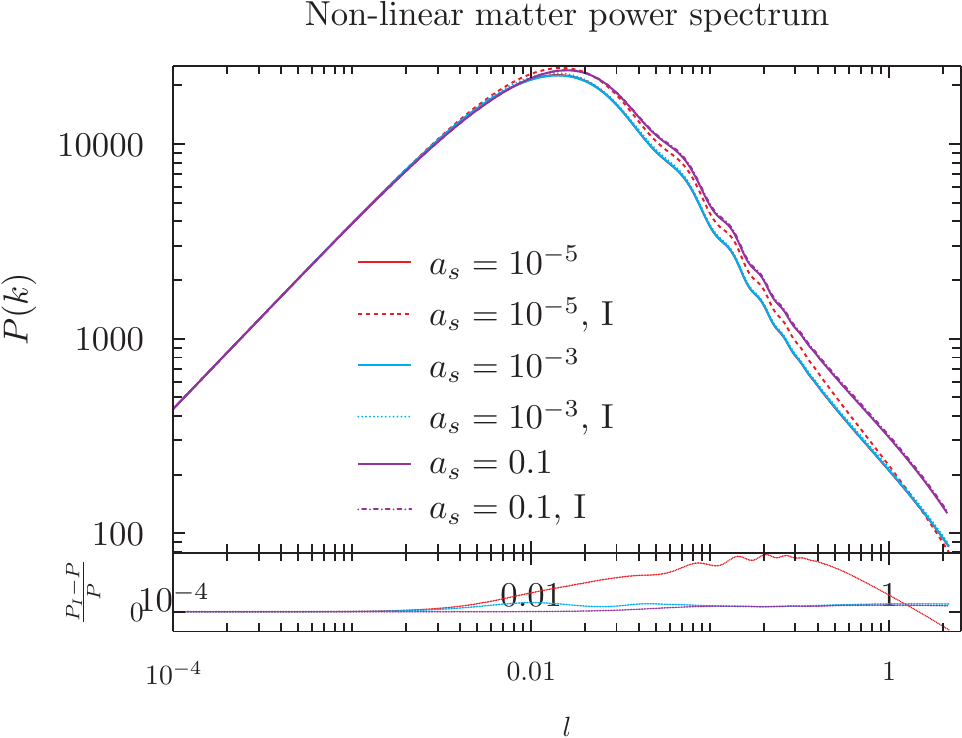}
}
\subfigure[]{
\includegraphics[scale = 0.7]{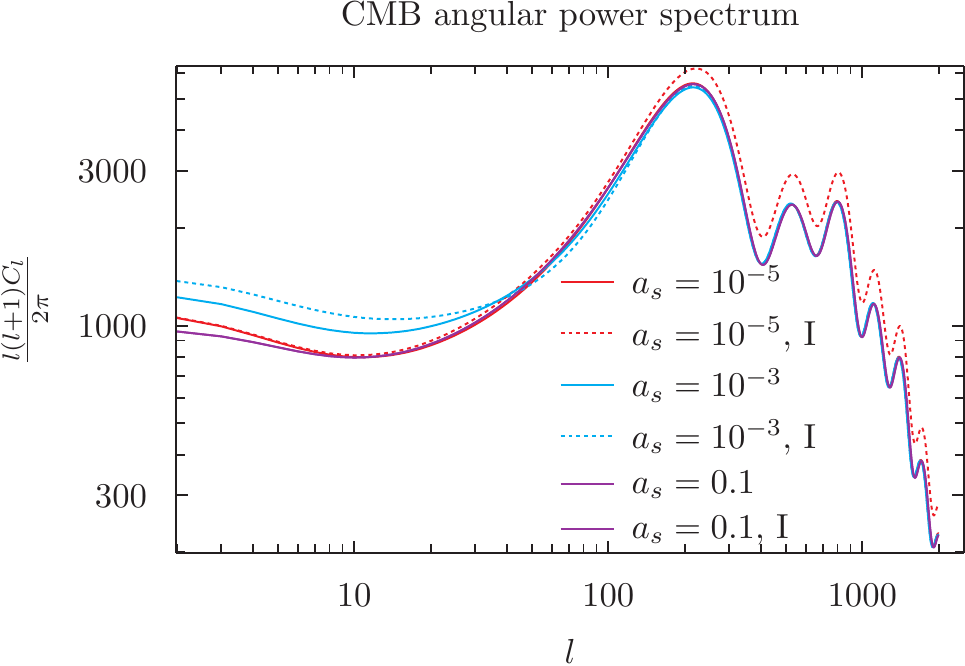}
}
\caption{Power spectrum and CMB angular power spectrum comparison for models with and without neutrino self-interactions after $a = a_{s}$, for variable values of $a_{s}$ and all other parameters kept fixed, with $m_{\nu} =0.31$eV and $B_{s} = 10^{10}$ and other parameters set in accordance with the standard $\Lambda$CDM model, {\em i.e.} $H_0 \approx 70$km/s/Mpc, $\Omega_{c} \approx 0.3$ and $\Omega_{\Lambda}\approx 0.7$. Legends with an ``I" indicate models where interactions are turned on at $a = a_s$.}
\label{fig:interact}
\end{figure}

The Vlasov equation describes the evolution of the neutrino distribution function when neutrinos are collisionless. When neutrinos are strongly self-interacting, the right-hand side of the Boltzmann equation will be non-vanishing. In general, the behavior of the neutrino distribution function depends on the specific form of this right-hand side. However, the main effect of having {\em strong} neutrino self-interactions is that the anisotropic stress of the neutrinos vanishes and the neutrinos effectively behave as a perfect fluid, corresponding to the multipoles $\Psi_l$ vanishing for $l\geq 2$ (see {\em e.g.} \cite{heirarchy_to_delta} for a derivation of this result or \cite{Pastor} for a qualitative discussion).
\newline\indent
Without anisotropic stress, the neutrinos will no longer free-stream but will instead cluster, leading to an increased power in the neutrino transfer function as illustrated in figure \ref{fig:transfer_nu}. The figure shows an increase in power on small scales for all the three shown cases. The increase in power is seen only on small scales because these correspond to the scales where free-streaming suppresses growth when the neutrinos are collisionless.
\newline\indent
Note that the relative transfer functions have a turnover at relatively high $k$. This happens because as soon as the mass is switched on, the transfer function grows very rapidly in both cases. This suppresses the relative transfer function simply because of the larger denominator.
\newline\newline
Overall, if a strong neutrino self-interaction is instantaneously turned on when $a = a_{s}$ by setting all $\Psi_l = 0$ for $l\geq 2$, the power spectrum will respond with an increased power on small scales. Similarly, the neutrino fluid will contribute to the acoustic oscillations and hence an increase in the CMB angular power spectrum occurs, primarily on scales that only enter the horizon {\em after} the neutrinos have begun interacting. These effects are visible in the power spectra and CMB angular power spectra of figure \ref{fig:interact}.

\FloatBarrier

\section{Cosmological models and data}\label{sec:data}
We use the standard $\Lambda$CDM model as the basis for our study but add massive neutrinos and self-interactions in four different combinations. In model 1 (M1) we add neutrinos which become massive when the scale factor reaches a specific, arbitrary value, $a_s\leq1$. In model 2 (M2) we in addition introduce strong self-interactions of the neutrinos once $a = a_s$ is reached. While the transition from $m_{\nu}(a) = 0$ to $m_\nu(a) = m_{\nu}$ can occur at different rates depending on $B_s$ (see equation (\ref{eq:mass})), the transition to strong self-interaction occurs instantaneously. Models 3 and 4 (M3 and M4) are equivalent to M1 and M2 except that $a_s$ is no longer arbitrary. Instead, $a_s$ is fixed so that it corresponds to the time when the neutrino temperature drops to the temperature corresponding to the present day value of the neutrino mass. In all four models, the neutrino masses are assumed degenerate.
\newline\indent
In addition to studying the above four models, a reference model is also studied for comparison. The reference model is the $\Lambda$CDM model extended with ``regular" massive neutrinos, {\em i.e.} neutrinos which are massive at all times with constant value of their mass.
\newline\newline
As illustrated in the previous section, neutrino masses and strong neutrino self-interactions can significantly affect the CMB and the growth of structure. It is therefore reasonable to use CMB data and some measure of structure formation data to constrain the models (see {\em e.g.} \cite{data} for detailed discussions of how neutrinos affect CMB and BAO observations). The most robust structure formation data is given by BAO data which at the same time contains most of the information needed to constrain standard model parameters and neutrino masses (see {\em e.g.} \cite{useBAO}). BAO data will therefore be used here as the data which contains information on the growth of structure.
\newline\indent
As discussed in {\em e.g.} \cite{data,cmbH0degenerate}, $H_0$ and the sum of neutrino masses are degenerate when considered with CMB data alone. It is therefore standard to include the HST estimate of $H_0$ when constraining neutrino parameters with cosmology. However, the $\sim 3\sigma$ disagreement between $H_0$ estimates from Planck and HST (see {\em e.g.} \cite{Riess:2016jrr}) could indicate that the local $H_0$ measurement is biased {\em e.g.} due to the local density distribution or late-time emergence of curvature (see {\em e.g.} \cite{HubbleTrouble} regarding the former and \cite{emergence} regarding the latter). We therefore choose not to use HST data here. We do however include supernova data in order to increase the impact of background effects. As is clear from the results of the next section, the supernova data has virtually no effect on parameter constraints though. This is not too surprising as the BAO data already incorporates background effects.
\newline\newline
In summary, using the publicly available CosmoMC package \cite{Lewis:2002ah}, the cosmological models are analyzed in relation to observational data, including CMB, BAO and supernovae. The CMB data includes the Planck 2015 \cite{Planck} TT and combined low-$l$ ($2\leq l\leq 29$) TT, EE, BB and TE likelihoods. The BAO data comes from SDSS DR11 \cite{sdss} and the 6dF survey \cite{6df}. The supernova data is that from SNLS \cite{Conley:2011ku}, union2 \cite{union2}, and the JLA project \cite{Betoule:2014frx}.
\newline\newline
 The cosmological parameters constrained in the analyses are
\begin{equation}
\left\lbrace\sum m_{\nu}, \log(a_s), \log(B_s), \Omega_{c}h^2, \Omega_{b}h^2,\theta_s, n_s, \tau, \ln(10^{10}A_s) \right\rbrace ,
\end{equation}
with the following variations: In M3 and M4, $a_s$ is not an independents parameter but is determined by $m_{\nu}$, and in the reference model, $a_s$ and $B_s$ are both irrelevant. The parameters $\theta_s$, $\tau$, $n_s$ and $A_s$ are, respectively, the angular size to the horizon, the optical depth to reionization, and the spectral index and amplitude of the scalar fluctuations induced by inflation. $\Omega_c,\Omega_b$ are the present time density parameters of CDM and baryonic matter while $h$ is the reduced Hubble constant. 

\section{Results}\label{sec:Results}
\begin{table}[!htb]
\resizebox{\columnwidth}{!}{
\centering
\begin{tabular}{c c c c c c c c c c c}
\hline\hline 
\Tstrut
Model & $\sum m_{\nu}, \left[ \text{eV}\right] $ & $\log(a_s)$& $\log(B_s)$  & $\Omega_{b}h^2$ & $\Omega_{c}h^2$ & $H_0, \left[\text{km/s/Mpc} \right] $& $n_s$& $\sigma_8$& $\tau$&  $10^9 A_s$\\
\hline \Tstrut
ref, CMB & $0.03679^{+0.73}$  &$\div$  &$\div$ & $0.02241^{+0.00020}_{-0.00084}$  & $0.1191^{+0.0059}_{-0.0032}$  & $67.97^{+1.5}_{-8.2}$  & $0.9684^{+0.00082}_{-0.019}$ & $0.8471^{+0.017}_{-0.16}$ & $0.09486^{+0.024}_{-0.052}$ & $2.263^{+0.11}_{-0.21}$ \\
ref, CMB+BAO & $0.05066^{+0.16}$ & $\div$ & $\div$& $0.02236^{+0.00032}_{-0.00047}$  & $0.1201^{+0.091}_{-0.0042}$ & $67.46^{+1.3}_{-1.1}$ & $0.9647^{+0.013}_{-0.0060}$ & $0.8273^{+0.036}_{-0.041}$ & $0.07025^{+0.049}_{-0.024}$ & $2.161^{+0.22}_{-0.10}$ \\
ref, CMB+BAO+Sn & $0.07237^{+0.14}$ & $\div$ & $\div$ & $0.02218^{+0.00052}_{-0.00028}$ & $0.1197^{+0.0018}_{-0.0039}$ & $67.75^{+1.1}_{-1.5}$ & $0.9646^{+0.013}_{-0.0063}$ & $0.8383^{+0.026}_{-0.054}$ & $0.07237^{+0.048}_{-0.027}$ & $2.178^{+0.20}_{-0.12}$ \\
M1, CMB & $1.288^{+0.26}$ & $-1.325$ & $-0.4425$ & $0.02228^{+0.00042}_{-0.00055}$ & $0.1191^{+0.00054}_{-0.0036}$ & $61.03^{+8.0}_{-2.1}$ & $0.9665^{+0.011}_{-0.014}$ & $0.7185^{+0.14}_{-0.030}$ & $0.08233^{+0.034}_{-0.040}$ & $2.212^{+0.15}_{-0.16}$ \\
M1, CMB+BAO & $0.3466^{+0.27}$  & $-1.241$ & $2.205$ & $0.02222^{+0.00049}_{-0.00034}$ & $0.1186^{+0.0031}_{-0.0030}$ & $65.79^{+3.1}_{-0.38}$ & $0.9693^{+0.0086}_{-0.017}$ & $0.7881^{+0.075}_{-0.015}$ & $0.08251^{+0.035}_{-0.038}$ & $2.207^{+0.16}_{-0.15}$ \\
M1, CMB+BAO+Sn  & $1.043^{+0.54}$  & $-0.9492$ &$-0.7584$ & $0.02237^{+0.00034}_{-0.00047}$ & $0.1180^{+0.0036}_{-0.0023}$  & $66.99^{+1.9}_{-0.77}$ & $0.9692^{+0.0087}_{-0.011}$ & $0.7821^{+0.081}_{-0.0052}$ & $0.08418^{+0.034}_{-0.040}$ & $2.211^{+0.16}_{-1.2}$ \\
M2, CMB  & $0.04498^{+0.63}$  & $-1.867$ & $-0.6071$ & $0.02232^{+0.00040}_{-0.00063}$ & $0.1179^{+0.0053}_{-0.0048}$  & $68.37^{+2.3}_{-5.8}$ & $0.9645^{+0.0093}_{-0.018}$ & $0.8378^{+0.027}_{-0.094}$ & $0.08082^{+0.088}_{-0.040}$ & $2.182^{+0.84}_{-0.13}$ \\
M2, CMB+BAO & $0.1104^{+0.53}$ & $-0.8624_{-1.4}$ & $3.687$  & $0.02222^{+0.00045}_{-0.00037}$ & $0.1178^{+0.0031}_{-0.0031}$ & $67.92^{+1.1}_{-1.7}$  & $0.9645^{+0.0069}_{-0.013}$ & $0.8373^{+0.025}_{-0.060}$ & $0.09087^{+0.023}_{-0.048}$  & $2.222^{+0.11}_{-0.20}$   \\
M2, CMB+BAO+Sn  & $0.3352^{+0.31}$ & $-0.5360_{-1.7}$ & $3.415$ & $0.02236^{+0.00031}_{-0.00050}$ & $0.1169^{+0.0040}_{-0.0021}$ & $67.84^{+1.2}_{-1.6}$ & $0.9648^{+0.0065}_{-0.013}$ & $0.8163^{+0.045}_{-0.040}$ & $0.08607^{+0.027}_{-0.043}$ & $2.210^{+0.12}_{-0.18}$\\
M3, CMB  & $0.007135^{+0.50}$ & $\div$  & $1.029$ & $0.02235^{+0.00029}_{-0.00065}$ & $0.1193^{+0.0058}_{-0.0035}$ & $68.10^{+1.4}_{-6.6}$  & $0.9682^{+0.0088}_{-0.018}$ & $0.8516^{+0.012}_{-0.12}$ & $0.08965^{+0.029}_{-0.047}$ & $2.252^{+0.12}_{-0.20}$\\
M3, CMB+BAO  & $0.008211^{+0.21}$& $\div$ & $1.011$ & $0.02236^{+0.00035}_{-0.00047}$ &$0.1195^{+0.0020}_{-0.0037}$ & $68.10^{+0.79}_{-1.8}$ & $0.9643^{+0.013}_{-0.0057}$ & $0.8492^{+0.014}_{-0.065}$ & $0.08670^{+0.032}_{-0.042}$ & $2.241^{+0.13}_{-0.18}$ \\
M3, CMB+BAO+Sn  & $0.03563^{+0.18}$ & $\div$ & $0.5051$ & $0.02226^{+0.00044}_{-0.00035}$ & $0.1206^{+0.00090}_{-0.0048}$& $67.35^{+1.5}_{-1.1}$ & $0.9644^{+0.013}_{-0.0056} $ & $0.8406^{+0.022}_{-0.055}$ & $0.07657^{+0.042}_{-0.031}$ & $2.202^{+0.17}_{-0.15}$ \\
M4, CMB  & $0.01559^{+0.15}$ & $\div $& $-0.5847$ & $0.02232^{+0.00039}_{-0.00051}$& $0.1166^{+0.0057}_{-0.0029}$ & $68.78^{+1.7}_{-3.9}$ & $0.9656^{+0.0080}_{-0.017}$ & $0.8336^{+0.040}_{-0.068}$& $0.08455^{+0.032}_{-0.042}$ & $2.196^{+0.14}_{-0.17}$ \\
M4, CMB+BAO  & $0.07316^{+0.077}$  & $\div$ & $2.175$& $ 0.02223^{+0.00034}_{-0.00045}$& $0.1185^{+0.0027}_{-0.0030}$ &$67.63^{+1.6}_{-1.4}$  & $0.9624^{+0.0084}_{-0.011}$ &$0.8361^{+0.035}_{-0.057}1$ &$0.08375^{+0.030}_{-0.040}$ & $2.202^{+0.13}_{-0.17}$ \\
M4, CMB+BAO+Sn  & $0.01919^{+0.14}$ & $\div$ &$2.464$ &$0.02226^{+0.00040}_{-0.00041}$ &$0.1188^{+0.0020}_{-0.0035}$ & $68.02^{+1.1}_{-1.7}$ &$0.9578^{+0.012}_{-0.0062}$  & $0.8374^{+0.033}_{-0.059}$ & $0.07069^{+0.043}_{-0.028}$ & $2.154^{+0.18}_{-0.13}$ \\
\hline
\end{tabular} }
\caption{Best fits and 95 \% confidence intervals for the reference model (ref), and models M1, M2, M3 and M4, using the data described in section \ref{sec:data}. Higher or lower limits are not indicated when they correspond to the prior limits. This is the case for the lower limit of $\sum m_{\nu}$ which is simply zero. It is also often the case for $\log_{10}(a_s)$ which has the prior lower limit of $-5$ and upper limit of zero, and for $\log_{10}(B_s)$ which has been given the prior lower limit of $-1$ and upper limit of $4$.}
\label{table:BestFit}
\end{table}

\begin{figure}[htb!]
\centering
\includegraphics[scale = 0.6]{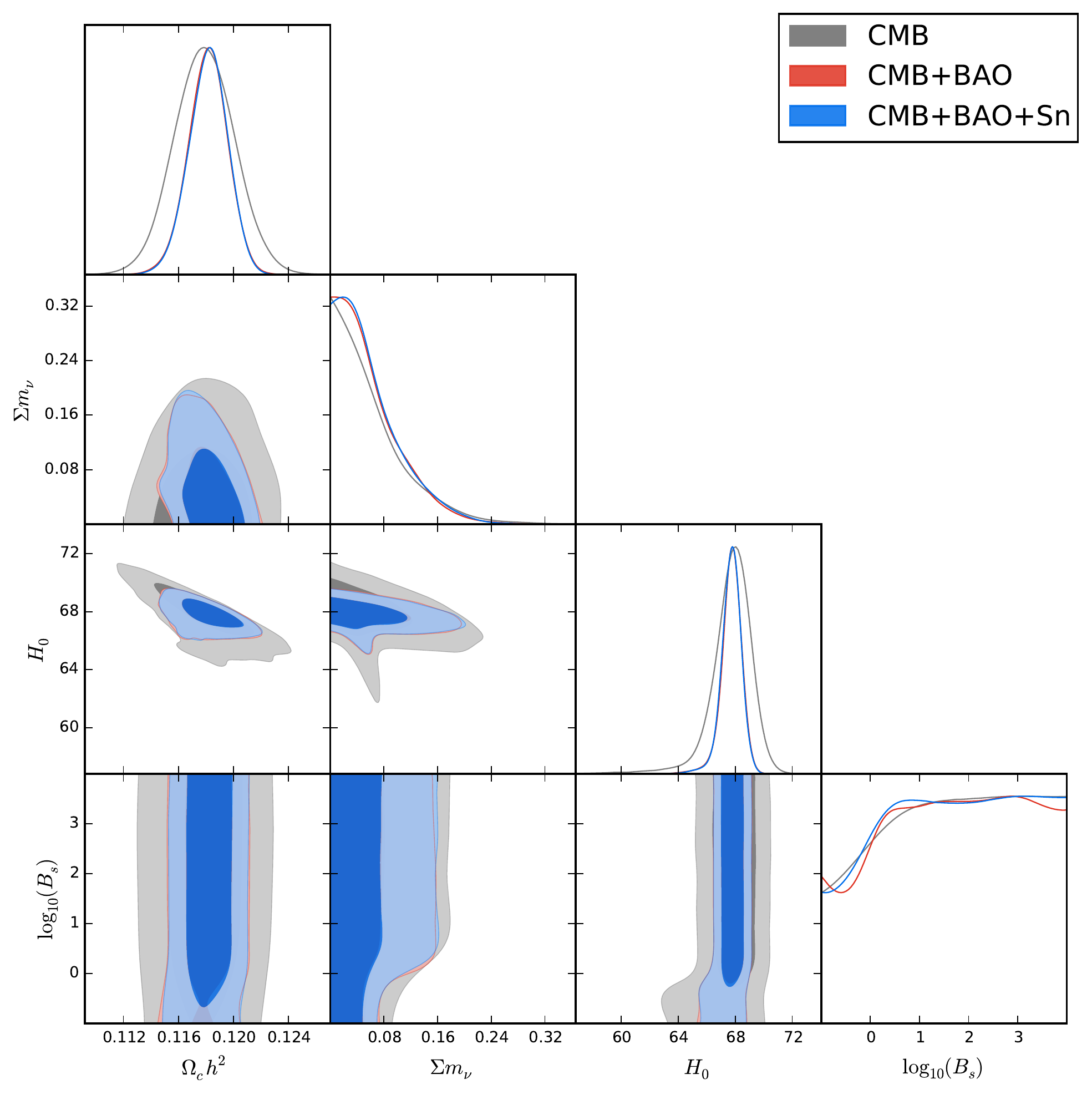}
\caption{Triangle plot showing the marginalized probability distribution functions and 2D parameter spaces for $\Sigma m_{\nu}$, $H_0$, $\Omega_c h^2$ and $\log_{10}(B_s)$. The figures were computed using GetDist with data from CosmoMC runs of M4.}
\label{fig:m4}
\end{figure}

\begin{figure}[htb!]
\centering
\includegraphics[scale = 0.6]{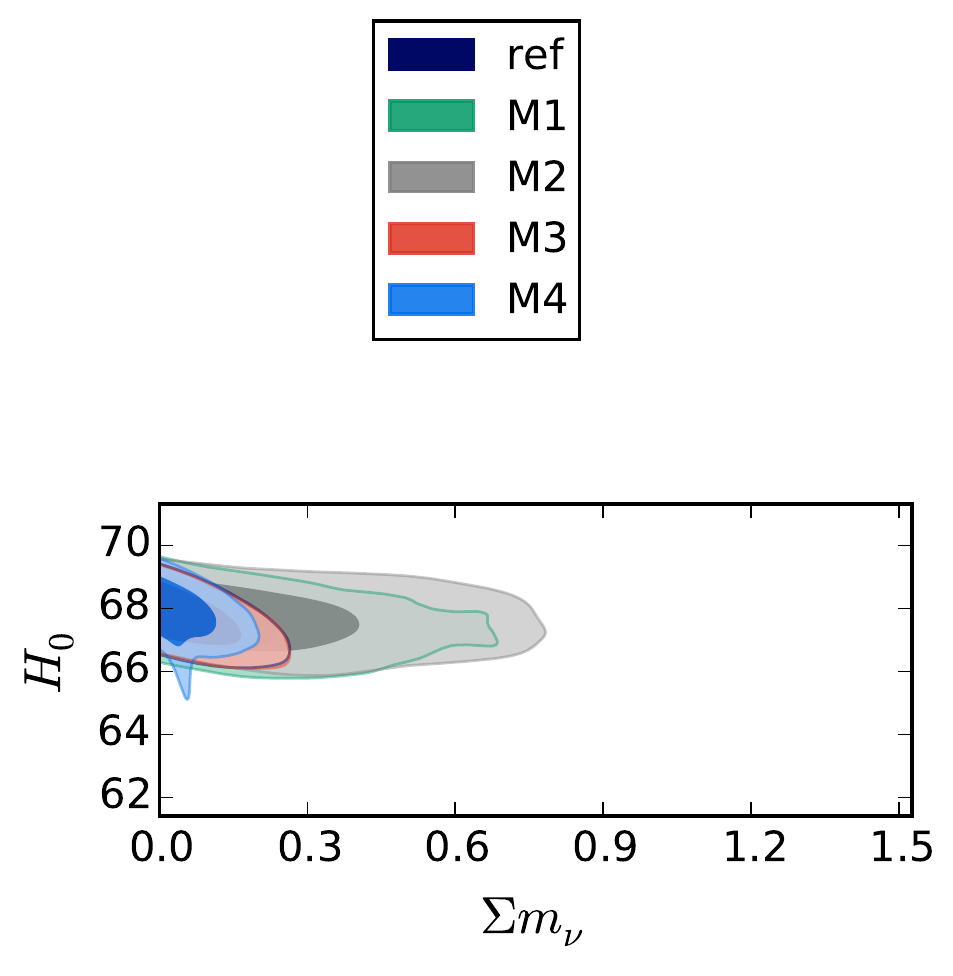}
\caption{Combined allowed 2D parameter space for $\Sigma m_{\nu}$ and $H_0$ for all five studied models. The figures were computed using GetDist with data from CosmoMC runs using the both CMB, BAO and supernova data.}
\label{fig:2D}
\end{figure}

In this section, we present and discuss the results obtained from comparing the cosmological models with data. In particular, the best fit and $95\%$ confidence intervals for the model parameters obtained from running CosmoMC with the four models as well as the reference model are shown in table \ref{table:BestFit}. In addition, triangle plots for selected parameters for each model are shown.
\newline\newline
Table \ref{table:BestFit} shows that the best fit values of $\Omega_b, \Omega_c, n_s$ and $A_s$ are very similar for all models, while M1 has a slightly smaller amount of structure formation on small scales than the other models, with $\sigma_8\sim 0.78$ instead of $\sigma_8\sim 0.81-0.84$. This happens because the best-fit neutrino mass is higher in M1 and therefore the same value of $A_s$ leads to a lower value of $\sigma_8$ because of neutrino suppression of structure growth. In addition, $H_0$ takes on noticeably lower values in some of the models compared to the reference model and the acceptable upper limit on $\sum m_{\nu}$ varies notably amongst the models. These features will be discussed below.
\newline\newline
\begin{figure}[htb!]
\centering
\includegraphics[scale = 0.6]{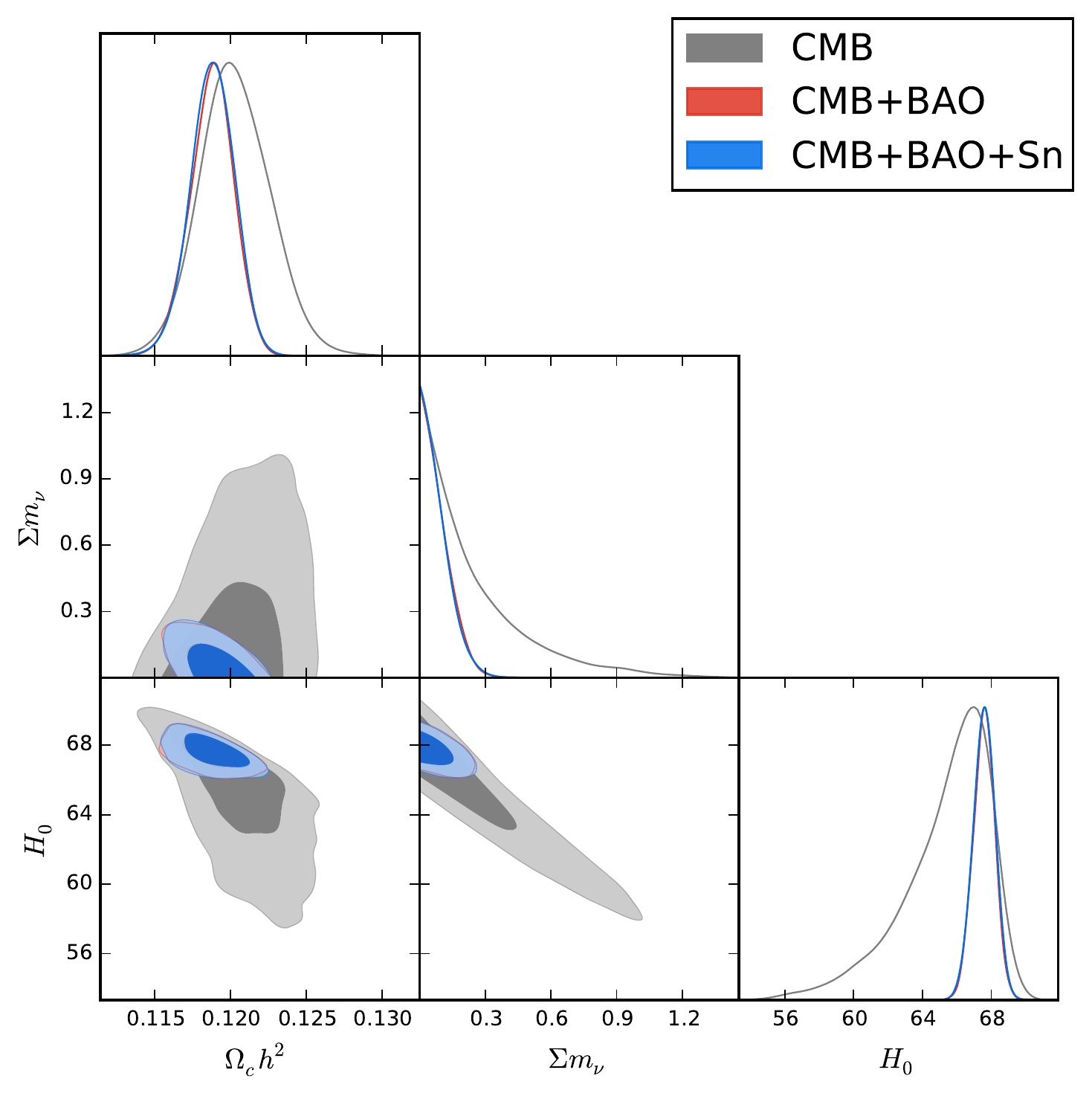}
\caption{Triangle plot showing the marginalized probability distribution functions and 2D parameter spaces for $\Sigma m_{\nu}$, $H_0$ and $\Omega_c h^2$. The figures were computed using GetDist with data from CosmoMC runs of the reference model.}
\label{fig:ref}
\end{figure}
Figure \ref{fig:ref} shows a triangle plot for the reference model with a few selected model parameters. The plots are in agreement with those found elsewhere. Adding BAO data leads to much more strict constraints on the upper limit of the sum of neutrino masses than the CMB alone does. As expected, adding supernova data barely changes the constraints compared to only using CMB and BAO data.
\newline\newline
\begin{figure}[htb!]
\centering
\includegraphics[scale = 0.6]{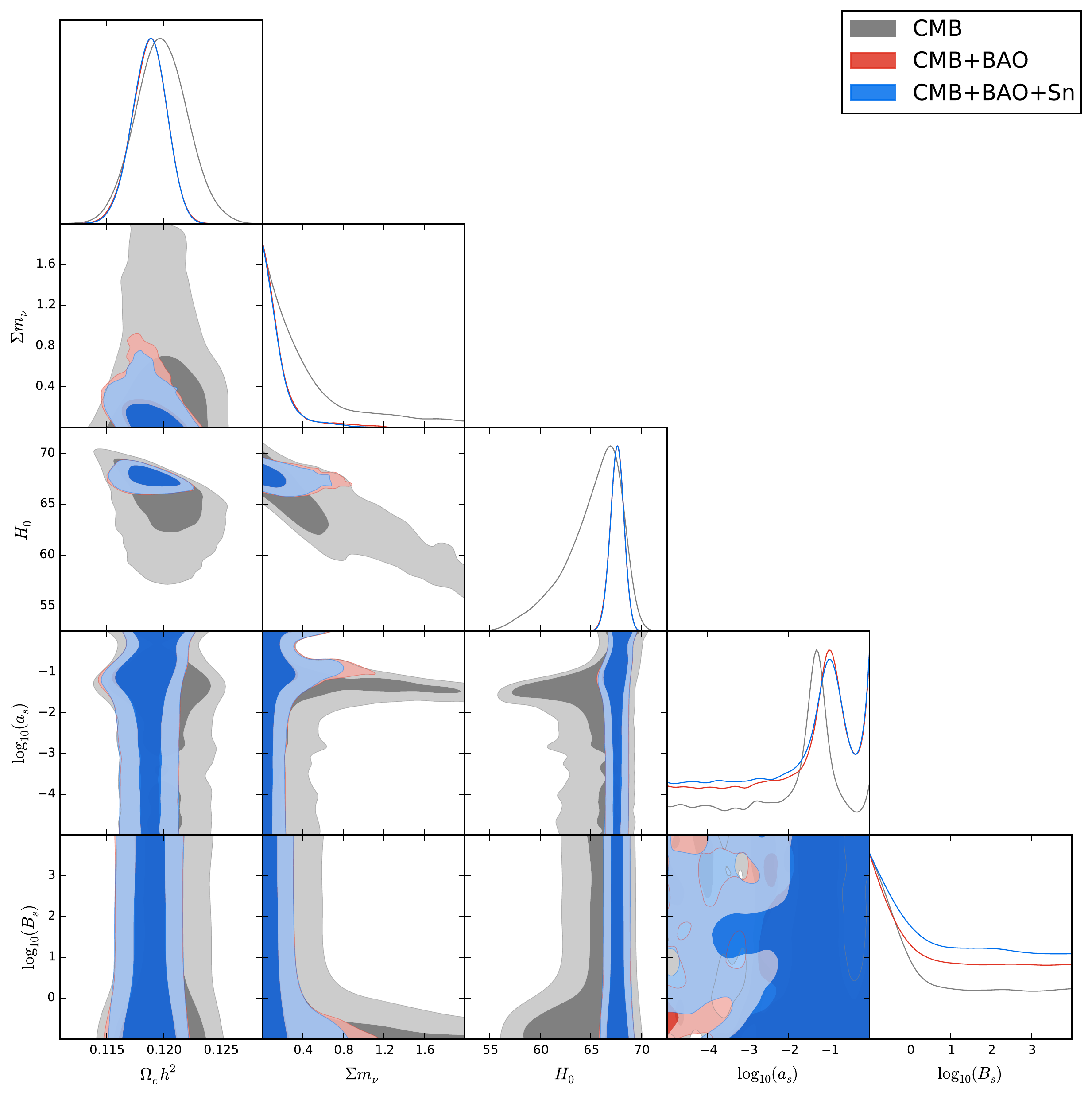}
\caption{Triangle plot showing the marginalized probability distribution functions and 2D parameter spaces for $\Sigma m_{\nu}$, $H_0$, $\Omega_c h^2$, $\log_{10}(a_s)$ and $\log_{10}(B_s)$. The figures were computed using GetDist with data from CosmoMC runs of M1.}
\label{fig:m1}
\end{figure}
For M1 (figure \ref{fig:m1}), we see that as long as only CMB data is used, the sensitivity to the late time behavior of the neutrino mass is very limited. This means that a very high neutrino mass is possible provided that $a_s$ is tuned accordingly. However, once BAO data is added, the sensitivity to $m_{\nu}$ increases significantly and it is no longer possible to hide a large present day neutrino mass by switching it on very late in the evolution of the Universe. For the CMB data alone, the constraints on $\sum m_\nu$ looks somewhat as though the chains have not converged. In reality, however, the somewhat unconverged appearance is due to the fact that high values of $a_s$ are permitted; this renders the neutrinos effectively massless during almost the entire history of the Universe which again means that the sensitivity to $m_\nu$ is limited.
\newline\indent
The sensitivity to $B_s$ and $a_s$ is quite limited, although the marginalized probability distributions indicate a slight preference to larger values of $a_s$ and low values of $B_s$. Figure \ref{fig:cmb_special} shows that $B_s$ and $a_s$ affect the same part of the CMB angular power spectrum, namely primarily the low-$l$ part. The overall effect of the two parameters on the CMB angular power spectrum is to raise and lower it, depending on their particular values. Unfortunately, the effects of the two parameters do not simply counter each other in a simple way such that large values of one would consistently permit large values of the other or the like. Instead, the effects of the two parameters counter each other in a quite complicated way, leading to a quite messy 2D plot of these two parameters in figure \ref{fig:m1}.
\newline\indent
Regarding figure \ref{fig:m1} we lastly note that in the case of combined CMB, BAO and supernovae data, the best fit point found is outside the formal 95\% confidence region. This presumably happens because the allowed region at high $\sum m_{\nu}$ is quite narrow and requires very specific values of $a_s$. This means that it becomes disfavored by the volume effect in the likelihood integral and that the formal allowed region is shifted to lower values of $\sum m_{\nu}$ even though very good fits remain at large values of $\sum m_{\nu}$.
\newline\newline
\begin{figure}[htb!]
\centering
\includegraphics[scale = 0.6]{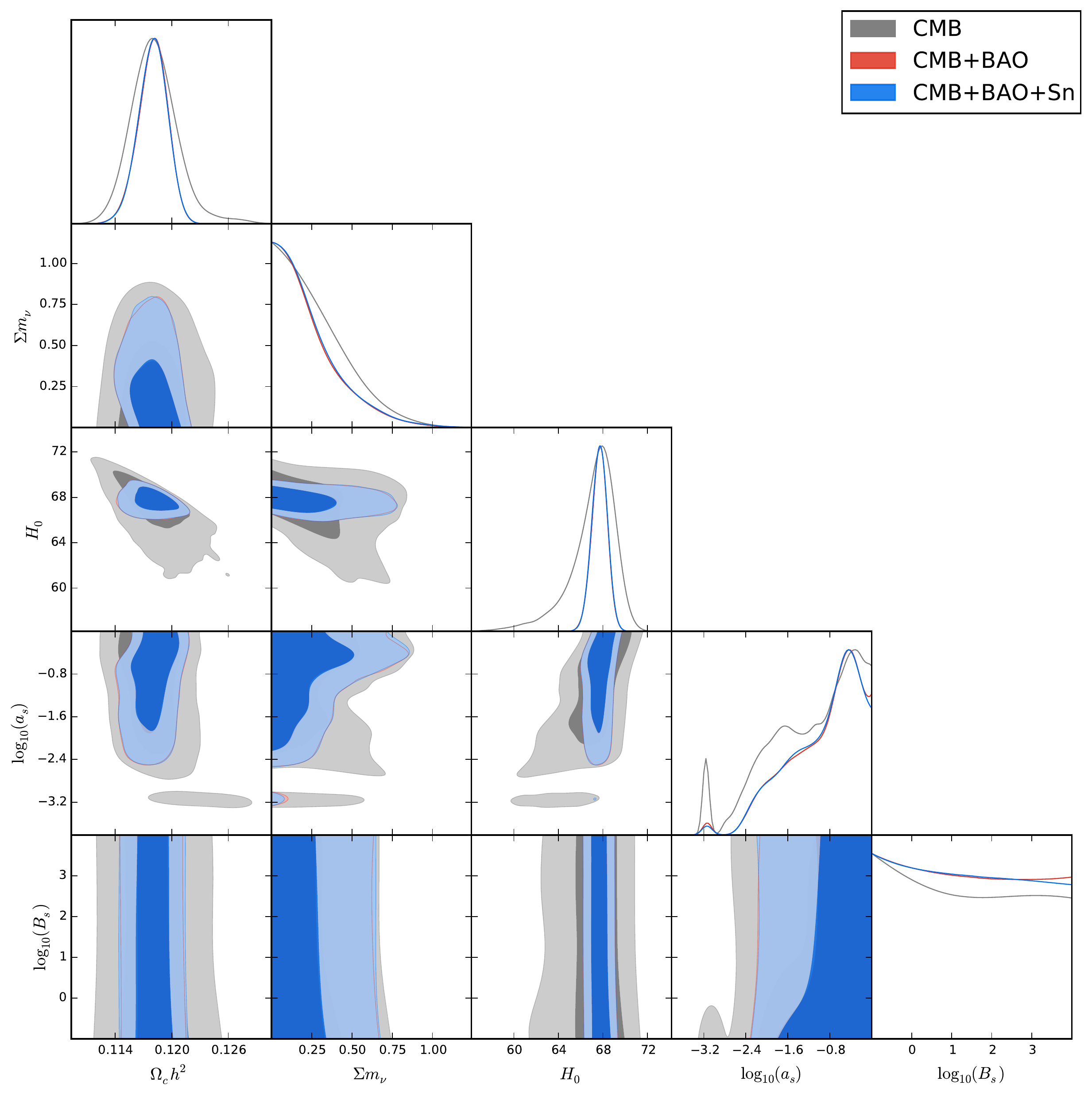}
\caption{Triangle plot showing the marginalized probability distribution functions and 2D parameter spaces for $\Sigma m_{\nu}$, $H_0$, $\Omega_c h^2$, $\log_{10}(a_s)$ and $\log_{10}(B_s)$. The figures were computed using GetDist with data from CosmoMC runs of M2.}
\label{fig:m2}
\end{figure}
In M2 (figure \ref{fig:m2}), the effect of switching on the neutrino mass deviates from that seen for M1. This is because anisotropic stress vanishes at the point where the mass is switched on since this event is now accompanied with the onset of strong neutrino self-interactions. The CMB is quite sensitive to the presence of anisotropic stress in the neutrino component and therefore the CMB sensitivity to $a_s$ is now much better than in M1. The sensitivity to $B_s$ is even more limited than for M1. This is presumably because the sensitivity to $B_s$ is most prominent for $a_s \sim 10^{-3}$ while the data now clearly prefers a larger value of $a_s$ (disregarding the small peak around $\log_{10}(a_s)\approx -3.2$ which is discussed below).
\newline\indent
Overall, CMB data prefers free-streaming neutrinos and the allowed parameter space is large for high values of $a_s$. However, there is a small peak in the marginalized probability distribution function of $\log_{10}(a_s)$ around a value slightly below $a_s\sim 10^{-3}$, {\em i.e.} for $a_s$ around the time of last scattering. This leads to small islands in the allowed parameter spaces of other parameters. The peak is most prominent in the case where only CMB data is considered, indicating that it arises because effects of the neutrino mass generation on the CMB angular power spectrum is canceled by the onset of strong interactions. A similar, but reverse, result was found in \cite{Lancaster:2017ksf} where neutrinos are strongly interacting at early epochs but decouple later. This leads to a peak in the posterior probability distribution of the coupling constant at high values. A similar result was later confirmed in \cite{Oldengott:2017fhy}.
\newline\indent
Figure \ref{fig:m2} also shows that the strong self-interactions lead to lower values of $H_0$. This again leads to even stronger tension with the local measurements of $H_0$ than the standard model since local measurements favor values around $H_0 \sim 72-73 \, {\rm km} \, {\rm s}^{-1} \, {\rm Mpc}^{-1}$ \cite{Riess:2016jrr}.
\newline\indent
Lastly, it is noted that $a_s = 1$ seems to be permitted in both M1 and M2, implying that all the considered data sets are consistent with neutrinos that are massless until present time.
\newline\newline
\begin{figure}[htb!]
\centering
\includegraphics[scale = 0.6]{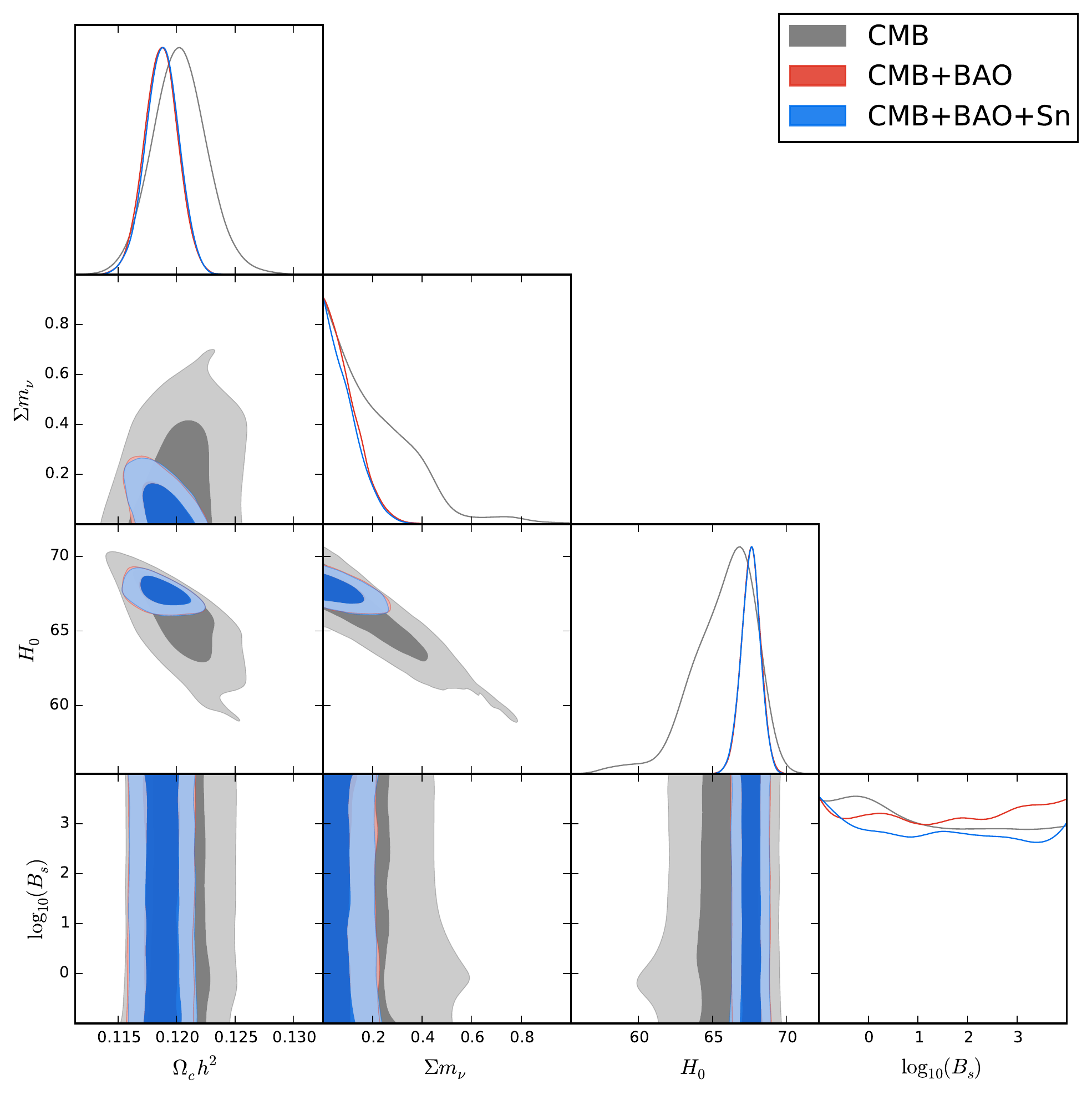}
\caption{Triangle plot showing the marginalized probability distribution functions and 2D parameter spaces for $\Sigma m_{\nu}$, $H_0$, $\Omega_c h^2$ and $\log_{10}(B_s)$. The figures were computed using GetDist with data from CosmoMC runs of M3.}
\label{fig:m3}
\end{figure}
Theoretically, M3 (figure \ref{fig:m3}) is clearly more restrictive than M1 and indeed we see that the large values of $\sum m_{\nu}$ allowed for both M1 and M2 by the combined data sets are not viable here. This happens because it in M3 is no longer possible to shift $a_s$ to high values while keeping $\sum m_{\nu}$ large (because $a_s \propto 1/\sum m_{\nu}$). This renders the allowed parameter intervals of M3 very similar to those of the reference model.
\newline\newline
M4 (figure \ref{fig:m4}) is the most restrictive of the four cases we study and its phenomenology corresponds roughly to that of \cite{dvali}. Since this model fixes the relation between the present day neutrino mass and the point where neutrinos become strongly interacting, the island of allowed parameter space around $a_s \sim 10^{-3}$ seen for M2 is no longer allowed (as can be seen from figure \ref{fig:m2}, this island requires $\sum m_{\nu}$ to be much smaller than $T(a=a_s)$).
\newline\indent
Like M1, M4 has a marginal sensitivity to the transition width, parameterized by $B_s$. Opposite to the case of M1, low values of $B_s$ are now slightly {\em dis}favored, implying that a faster transition is preferred.
\newline\indent
Lastly, it is noted that, similarly to the case of M2, for the case of CMB data only, a small region a low $H_0$ remains viable. The region is much less striking than in the case of M2 though.
\newline\newline
Figure \ref{fig:2D} shows the allowed 2D parameter space for $\Sigma m_{\nu}$ and $H_0$ for all five models. As seen, M1 and M2, where the neutrino mass is allowed to turn on at an arbitrary time (no later than the present), the neutrino mass is permitted to be much larger than in the reference model. Indeed, in these models, $\sum m_{\nu} \sim 0.6-0.8$eV is allowed even when both CMB, BAO and supernova data is used. Such a high value of the neutrino mass could make it accessible to direct measurement by the KATRIN experiment \cite{Osipowicz:2001sq}, which has a projected sensitivity to the effective electron neutrino mass around 0.2eV.
\newline\indent
M3 and M4 do not permit particularly large values of  $\Sigma m_{\nu}$. In fact, M4 which corresponds to the model in \cite{dvali} has a smaller range of allowed neutrino masses than the reference model does. The contributions of the reference model and M3 are almost indistinguishable in the figure.

Finally we note here that the phenomenological models studied here do not cover all possible models. For example the model of Dvali and Funcke contains the possibility that neutrinos pair annihilate to the light Goldstone bosons of the model. In this case the neutrinos and Goldstone bosons will effectively form a strongly interacting fluid (see e.g.\ \cite{Archidiacono:2014nda} for an example of this) . Such models can also be constrained using cosmological data, but we leave this for future study.

\FloatBarrier

\section{Conclusions}\label{sec:Conclusion}
In this work we have performed a detailed phenomenological study of the possibility that neutrino masses are generated dynamically at cosmologically late times. In order to be as general as possible we parameterized the neutrino mass in terms of its present day value, $m_{\nu}$, as well as two effective parameters, $a_s$ and $B_s$, describing the scale factor at the transition point and the width of the transition, respectively.
\newline\indent
We found that current cosmological data has little or no sensitivity to the parameter $B_s$ which indicates that at the current level of precision no further parameters are needed to adequately map models of dynamical mass generation. We also studied the possibility that neutrinos become strongly self-interacting at the time of mass generation, as is {\em e.g.} predicted in \cite{dvali}.
\newline\indent
As expected, we found that in general significantly higher values of the present day neutrino mass are allowed provided that the transition occurs sufficiently late. In such cases, a direct detection of a non-zero neutrino mass by the KATRIN experiment might be feasible \cite{Osipowicz:2001sq}, and the combination of cosmology with KATRIN data could allow for a direct mapping of a time-varying neutrino mass.
\newline\indent
In the case where the transition occurs at an energy scale corresponding to the present day neutrino mass, cosmological data puts a much stricter bound on the mass of the neutrino (and correspondingly on $a_s$). Again, this is not surprising since a large value of the neutrino mass requires the transition to happen early where both CMB and BAO data are sensitive to neutrino phenomenology.
\newline\indent
We furthermore find that models where neutrinos become strongly interacting at late times are viable for a relatively wide range in $a_s$. Intriguingly we also find that values of $a_s$ around the epoch of last scattering can be allowed by data. Models with strong self-interactions around recombination were also recently found to be allowed in the context of interactions mediated by a new massive vector boson \cite{Lancaster:2017ksf,Oldengott:2017fhy}. However, the models we study here are phenomenologically very different because in our case neutrinos become strongly self-interacting at late times while in the massive vector case the self-interactions freeze out at late times. In any case, models with strong neutrino self-interactions are very interesting and merit further study.

\section{Acknowledgments}
S. M. Koksbang thanks users of cosmocoffee for help with CAMB and CosmoMC, and Thomas Tram for help with CLASS. The numerical computations for this work were done using computing resources from the Center for Scientific Computing Aarhus.
We thank Lena Funcke for valuable comments.


\begin{thebibliography}{99}


\bibitem{kamiokande} The Super-Kamiokande Collaboration: Measurement of the flux and zenith-angle distribution of upward through-going muons by Super-Kamiokande, Phys.Rev.Lett.82:2644-2648,1999, arXiv:hep-ex/9812014v2 
\bibitem{solarSNO} SNO Collaboration: Direct Evidence for Neutrino Flavor Transformation from Neutral-Current Interactions in the Sudbury Neutrino Observatory, Phys.Rev.Lett.89:011301,2002, arXiv:nucl-ex/0204008v2
\bibitem{KamLAND} KamLAND Collaboration: First Results from KamLAND: Evidence for Reactor Anti-Neutrino Disappearance, Phys.Rev.Lett.90:021802,2003, arXiv:hep-ex/0212021v1

\bibitem{Esteban:2016qun}
  I.~Esteban, M.~C.~Gonzalez-Garcia, M.~Maltoni, I.~Martinez-Soler and T.~Schwetz,
  JHEP {\bf 1701} (2017) 087
  doi:10.1007/JHEP01(2017)087
  [arXiv:1611.01514 [hep-ph]].
	
\bibitem{Kraus:2004zw}
  C.~Kraus {\it et al.},
  Eur.\ Phys.\ J.\ C {\bf 40} (2005) 447
  doi:10.1140/epjc/s2005-02139-7
  [hep-ex/0412056].
	
\bibitem{fukugita_yanagida} M. Fukugita and T. Yanagida: The Physics of Neutrinos and Applications to Astrophysics, Springer, Berlin, Germany (2003).


\bibitem{mass_review} S. F. King: Neutrino Mass Models, Rept.Prog.Phys. 67 (2004) 107-158, arXiv:hep-ph/0310204v2 
\bibitem{seesaw_study} Manfred Lindner, Tommy Ohlsson, Gerhart Seidl: See-saw Mechanisms for Dirac and Majorana Neutrino Masses, Phys.Rev. D65 (2002) 053014, arXiv:hep-ph/0109264v2 

\bibitem{Fardon:2003eh} R. Fardon, A. E. Nelson and N. Weiner: Dark energy from varying neutrinos, JCAP {\bf 0410} (2004) 005,  [astro-ph/0309800]

\bibitem{Afshordi:2005ym} N. Afshordi, M. Zaldarriaga and K. Kohri: On the stability of dark energy with mass-varying neutrinos,Phys.\ Rev.\ D {\bf 72} (2005) 065024,  doi:10.1103/PhysRevD.72.065024,  [astro-ph/0506663]
	
\bibitem{dvali} Gia Dvali, Lena Funcke: Small neutrino masses from gravitational $\theta$-term, Phys. Rev. D 93, 113002 (2016), arXiv:1602.03191v4 [hep-ph]
\bibitem{similar_to_dvali} Z. Chacko, Lawrence J. Hall, Steven J. Oliver, Maxim Perelstein: Late Time Neutrino Masses, the LSND Experiment and the Cosmic Microwave Background, Phys.Rev.Lett.94:111801,2005, arXiv:hep-ph/0405067v1 
\bibitem{a_little_similar} Z.Chacko, Lawrence J. Hall, Takemichi Okui, Steven J. Oliver: CMB Signals of Neutrino Mass Generation, Phys.Rev. D70 (2004) 085008, arXiv:hep-ph/0312267v1

\bibitem{Lewis:1999bs} A. Lewis, A. Challinor and A. Lasenby: Efficient computation of CMB anisotropies in closed FRW models, Astrophys.\ J.\  {\bf 538} (2000) 473, [astro-ph/9911177]
\bibitem{Lewis:2002ah} A. Lewis and S. Bridle: Cosmological parameters from CMB and other data: A Monte Carlo approach,  Phys.\ Rev.\ D {\bf 66} (2002) 103511, [astro-ph/0205436]
\bibitem{Blas:2011rf} D. Blas, J. Lesgourgeus and T. Tram: The Cosmic Linear Anisotropy Solving System (CLASS) II: Approximation schemes, JCAP {\bf 1107} (2011) 034, [arXiv:1104.2933 [astro-ph.CO]]

\bibitem{MaAndBert} Chung-Pei Ma, Edmund Bertschinger: Cosmological Perturbation Theory in the Synchronous and Conformal Newtonian Gauges, Astrophys.J. 455 (1995) 7-25, arXiv:astro-ph/9506072v1 
\bibitem{heirarchy_to_delta} Steen Hannestad: Structure formation with strongly interacting neutrinos - implications for the cosmological neutrino mass bound, JCAP 0502:011,2005, arXiv:astro-ph/0411475v3
\bibitem{Pastor} Julien Lesgourgues, Sergio Pastor: Massive neutrinos and cosmology, Phys.Rept. 429 (2006) 307-379, arXiv:astro-ph/0603494v2 
\bibitem{dodelson} Scott Dodelson: Modern Cosmology, Academic press, Elsevier, 2003

\bibitem{data} Sunny Vagnozzi et al.: Unveiling $\nu$ secrets with cosmological data: neutrino masses and mass hierarchy, arXiv:1701.08172v1 [astro-ph.CO]
\bibitem{useBAO} Jan Hamann et al.: Cosmological parameters from large scale structure - geometric versus shape information, 10.1088/1475-7516/2010/07/022, arXiv:1003.3999v2 [astro-ph.CO]
\bibitem{cmbH0degenerate} Elena Giusarma, Roland de Putter, Olga Mena: Testing standard and non-standard neutrino physics with cosmological data, 10.1103/PhysRevD.87.043515,arXiv:1211.2154v1 [astro-ph.CO] 
\bibitem{Riess:2016jrr} A. G. Riess et al.: A 2.4\% Determination of the Local Value of the Hubble Constant, Astrophys.\ J.\  {\bf 826} (2016) no.1, 56, arXiv:1604.01424 [astro-ph.CO]
\bibitem{HubbleTrouble} Antonio Enea Romano: Hubble trouble or Hubble bubble?, arXiv:1609.04081v5 [astro-ph.CO]
\bibitem{emergence} Krzysztof Bolejko: Emergence of spatial curvature, arXiv:1707.01800 [astro-ph.CO]
\bibitem{Planck} The Planck Collaboration: Planck 2015 results. XIII. Cosmologicalparameters, Astron. Astrophys. 594 (2016) A13, arXiv:1502.01589 [astro-ph.CO]
\bibitem{sdss} Lauren Anderson et al.:The clustering of galaxies in the SDSS-III Baryon Oscillation Spectroscopic Survey: Baryon Acoustic Oscillations in the Data Release 10 and 11 galaxy samples, DOI 10.1093/mnras/stu523, arXiv:1312.4877v2 [astro-ph.CO] 
\bibitem{6df} F. Beutler et al.: The 6dF Galaxy Survey: Baryon Acoustic Oscillations and the Local Hubble Constant, Mon. Not. Roy. Astron. Soc. 416 (2011) 3017, arXiv:1106.3366[astro-ph.CO]


\bibitem{Conley:2011ku} SNLS Collaboration: Supernova Constraints and Systematic Uncertainties from the First 3 Years of the Supernova Legacy Survey, Astrophys.\ J.\ Suppl.\  {\bf 192} (2011) 1,  doi:10.1088/0067-0049/192/1/1, [arXiv:1104.1443 [astro-ph.CO]]

\bibitem{union2} N. Suzuki et al. (The Supernova Cosmology Project): The Hubble Space Telescope Cluster Supernova Survey: V. Improving the Dark Energy Constraints Above z>1 and Building an Early-Type-Hosted Supernova Sample, Astrophyys.\ J.\ {\bf 746}, 1 (2012), arXiv:1105.3470v1 [astro-ph.CO] 

\bibitem{Betoule:2014frx} M. Betoule et al. (SDSS Collaboration): Improved cosmological constraints from a joint analysis of the SDSS-II and SNLS supernova samples, Astron.\ Astrophys.\  {\bf 568} (2014) A22, doi:10.1051/0004-6361/201423413, arXiv:1401.4064 [astro-ph.CO]


\bibitem{Archidiacono:2014nda}
  M.~Archidiacono, S.~Hannestad, R.~S.~Hansen and T.~Tram,
  Phys.\ Rev.\ D {\bf 91} (2015) no.6,  065021
  doi:10.1103/PhysRevD.91.065021
  [arXiv:1404.5915 [astro-ph.CO]].

\bibitem{Lancaster:2017ksf} L. Lancaster, F. Y. Cyr-Racine, L. Knox and Z. Pan: A tale of two modes: Neutrino free-streaming in the early universe,  arXiv:1704.06657 [astro-ph.CO]

\bibitem{Oldengott:2017fhy} I. M. Oldengott, T. Tram, C. Rampf and Y. Y. Y. Wong: Interacting neutrinos in cosmology: exact description and constraints, arXiv:1706.02123 [astro-ph.CO]
\bibitem{Osipowicz:2001sq} A. Osipowicz et al. [Katrin collaboration]: KATRIN: A Next generation tritium beta decay experiment with sub-eV sensitivity for the electron neutrino mass. Letter of intent, hep-ex/0109033




  


\end{thebibliography}
\end{document}